\documentclass[seceq]{ptptex}

\newcommand{\be}{\begin{equation}} 
\newcommand{\ee}{\end{equation}}
\newcommand{\bea}{\begin{eqnarray}}
\newcommand{\eea}{\end{eqnarray}}

\newcommand{\gapp}{\mathrel{\raise.3ex\hbox{$>$}\mkern-14mu
              \lower0.6ex\hbox{$\sim$}}}
\newcommand{\lapp}{\mathrel{\raise.3ex\hbox{$<$}\mkern-14mu
              \lower0.6ex\hbox{$\sim$}}}

\newcommand\lsim{\lesssim}
\newcommand\gsim{\gtrsim}

\newcommand\vev[1]{{\langle {#1} \rangle}}
\renewcommand\({\left(}
\renewcommand\){\right)}
\renewcommand\[{\left[}
\renewcommand\]{\right]}

\newcommand\eq[1]{Eq.~(\ref{#1})}
\newcommand\eqs[2]{Eqs.~(\ref{#1}) and (\ref{#2})}

\newcommand\eqst[2]{Eqs.~(\ref{#1})--(\ref{#2})}

\newcommand\eqreff[1]{(\ref{#1})}

\newcommand\pa{\partial}

\newcommand\mpl{M_{\rm P}}

\newcommand{\dlabel}[1]{\label{#1}}

\def\calp{{\cal P}}

\def\calpz{{\calp_\zeta}}

\newcommand\bfk{{\mathbf k}}

\newcommand\bfp{{\mathbf p}}

\newcommand\bfx{{\mathbf x}}

\newcommand\GeV{\,\mbox{GeV}}
\newcommand\MeV{\,\mbox{MeV}}

\newcommand\sub[1]{_{\rm #1}}
\newcommand\su[1]{^{\rm #1}}

\newcommand\mone{^{-1}}
\newcommand\mtwo{^{-2}}

\newcommand\mfour{^{-4}}

\newcommand\mhalf{^{-1/2}}
\newcommand\half{^{1/2}}

\newcommand\mquarter{^{-1/4}}
\newcommand\threehalf{^{3/2}}

\newcommand\mn{{\mu\nu}}

\newcommand{\Ai}{\mbox{Ai}}
\newcommand{\Bi}{\mbox{Bi}}

\newcommand{\dpnad}{\delta p\sub{nad}}

\newcommand{\svev}{\sub{nl}}

\title{Issues concerning the waterfall of hybrid inflation}

\author{
David H.\  \textsc{Lyth}%
}

\inst{Department of Physics, Lancaster University, Lancaster LA1 4YB, UK}

\abst{
We discuss the waterfall that ends hybrid inflation. Making some simplifying
assumptions, that may be  satisfied by GUT inflation models,
two issues are addressed. First, the procedure of keeping the
quantum  fluctuation of the waterfall field only in the regime where it can be
regarded as classical. Second, the contribution to the primordial curvature
perturbation that is generated during the waterfall. Because the waterfall
field is heavy during inflation, the spectrum of this contribution is strongly
blue and hence is negligible on cosmological scales.
}

\begin{document}

\maketitle

\section{Introduction}

The initial condition for the observable universe presumably is set by an early era 
of inflation \cite{book}. To generate the observed 
primordial curvature perturbation,   inflation should be almost exponential
while cosmological scales leave the horizon, corresponding to an almost constant Hubble
parameter $H(t)$. The simplest way of achieving that is to invoke slow-roll inflation,
where $H(\phi(t))$ depends only on a slowly varying inflaton field  $\phi$.

The  particular type of  slow-roll 
 inflation called hybrid inflation  invokes, in  addition to the inflaton, 
 a waterfall field $\chi$. 
The inflaton has zero vev while the waterfall has a nonzero vev.
Until inflation nears its end, $\chi$ is  fixed at the origin  by 
 a positive mass-squared $m^2(\phi(t))$, up to a vacuum fluctuation that is
dropped.  The displacement of $\chi$ from its vev is supposed to
generate most  of the inflationary potential. 

When
$\phi(t)$ falls  through  some critical value, $m^2(\phi)$ goes negative.
 Then, for wavenumbers below some maximum $k\sub{max}$,
the vacuum fluctuation $\chi_\bfk(t)$ becomes classical. 
Only the classical modes are kept. The spatially averaged field $\chi(t)$
grows with time, and eventually becomes equal to the vev of $\chi$.
This growth of $\chi$ is called the waterfall because it is supposed to 
happen quickly. 

In this article we consider  the waterfall era 
 without reference to a particular inflationary potential,
under some simplifying assumptions.
Two fundamental issues present themselves.

One of them is the  dropping of the vacuum fluctuation in the quantum regime.
 The motivation  
for this comes from the well-known fact that it 
 gives infinite contributions to the mean-square field, the 
energy density and the pressure. With a cutoff at momentum $\Lambda\sub{UV}$,
the  energy density is $\rho\sub{vluc}=
\Lambda^4\sub{UV}/16\pi^2$, which is widely
regarded as a contribution to the cosmological constant. In fact, 
 that  is not a viable interpretation, 
because the vacuum fluctuation contributions gives 
pressure $P\sub{vfluc} = \Lambda^4\sub{UV}/48\pi^2$, whereas the cosmological
constant has   $P_\Lambda=- \rho_\Lambda$. We will discuss some alternative
proposals to deal with the vacuum fluctuation, 
 but none are relevant for the waterfall  
and we stay with the procedure dropping the quantum regime.

Our second issue concerns the contribution $\zeta_\chi$ to
the primordial curvature perturbation $\zeta$, that is generated during
the waterfall.
 In accordance with the received
wisdom for a field that is heavy before the waterfall begins,
we find that the spectrum of this contribution goes like $k^3$.
On the horizon scale at the end of inflation, it might be big enough for
significant black hole formation which would constrain the parameters
of the hybrid inflation model, 
but on cosmological scales it will almost certainly be negligible.

The layout of the paper is as follows. In Section \ref{shybrid} we 
review hybrid inflation. In Section \ref{stach} we state our
assumptions and identify the part of parameter space in which they are
valid. In Section \ref{sgen} we consider the mode decomposition of the 
waterfall field and in Section \ref{sclass} the evolution of the classical
field. In Section \ref{sprim} we calculate the contribution of the  waterfall
to the curvature perturbation. In Section \ref{sgut} we show how our 
results can apply to GUT inflation, and we conclude in Section \ref{sconc}.

\section{Hybrid inflation}
\dlabel{shybrid}

We are concerned only with the simplest kind of slow-roll inflation,
which assumes Einstein gravity and an inflaton with the  canonical
kinetic term. The energy density  is  $\rho=3\mpl^2H^2$, where
 $\mpl=(8\pi G)\mhalf =2\times 10^{18}\GeV$ is the reduced 
Planck scale.
During slow-roll inflation the  following are good approximations:
\be
\rho=3\mpl^2H^2\simeq V(\phi),\qquad |\dot H|\ll H^2,\qquad
 3H\dot\phi\simeq - V'(\phi),\qquad
|\ddot\phi|\ll H|\dot\phi|
.\dlabel{hofv}
\ee
(Throughout this paper, an over-dot means differentiation with respect to 
$t$ while a prime means differentiation with respect to the displayed 
argument.)
 On scales leaving the horizon during slow-roll inflation,
  the perturbation $\delta\phi$
gives a time-independent contribution to the curvature perturbation,
whose spectrum is
\be
\calp_{\zeta_\phi}(k) = \( H^2/2\pi\dot\phi \)^2
,\dlabel{calpzetaphi} \ee
 with the right hand side evaluated at the epoch of horizon crossing
$k=aH$.

Hybrid inflation is a particular kind of slow-roll inflation. It 
was proposed  \cite{andreihybrid} as a way of solving
a worry about  axion cosmology \cite{myfirstaxion}, that  subsequently
was more or less laid to rest \cite{axionstring}. 
Hybrid inflation was soon found \cite{cllsw,gutinf,supernatural,runningmass} 
to be a powerful tool for model building especially in the context of
supersymmetry.\footnote
{More recently, there is interest in hybrid inflation with a non-canonical
kinetic term, in particular DBI inflation \cite{dbi}. The only essential
feature, needed to make sense of the hybrid inflation paradigm,
is that the potential dominates the energy density.}

 We will  adopt the  potential
\be
V(\phi,\chi) = \( V_0 + V(\phi) \) \nonumber 
+\frac12\( -m^2 +  g^2 \phi^2 \) \chi^2 + \frac14\lambda \chi^4
, \dlabel{fullpot} \ee
with $0< \lambda\ll 1$  and $g\ll 1$.
The second bracket  is $m^2(\phi)$, which goes negative
when $\phi$ falls below $m/g$.
During inflation  $\chi$ is taken to vanish
so that $V$ is given by the first bracket. The vev of $\phi$ vanishes
and we take $V(\phi)$ to vanish at the vev. The inflationary potential 
is supposed to be dominated by $V_0$ which means $V(\phi)\ll V_0$.
The requirements that $V$ and $\pa V/\pa \chi$ vanish in the vacuum
give the vev $\chi_0$ and the inflation scale $V_0$.
\be
\chi_0^2=m^2/\lambda,\qquad V_0=m^4/2\lambda
\dlabel{chiandv}, \ee 
leading to
\be
6 (H/m)^2 = (\chi_0/\mpl)^2
. \dlabel{hoverm} \ee

Minor variants of \eq{fullpot} would make little difference to our
analysis. The  interaction $g^2\phi^2\chi^2$ might be absent or suppressed,
so that  $\phi^2$ is  replaced by a higher power. 
 The term $\lambda\chi^4$ might be absent of suppressed, to be replaced by
a higher power. For our purpose, these  variants are  equivalent to allowing
(respectively) $g$ and $\lambda$ to be  many orders of magnitude below
unity. When making estimates, we will assume instead that these  parameters
are of order $10\mone$ to $10\mtwo$. Also, $\phi$ might have two or  more 
components that vary during inflation. Our 
 analysis will  apply to that case,  if at each instant
the field basis is chosen so that
only $\phi$  varies. 

More drastic modifications are also possible,
 including inverted hybrid inflation \cite{inverted}
where $\phi$ is increasing during inflation, and mutated/smooth hybrid
inflation \cite{smoothhybrid}
where the waterfall field varies during inflation.
Our analysis does not apply to those cases.

Taking $\chi(t)$ to be unperturbed during the waterfall, the field
equations are
\bea
\ddot \phi + 3H\dot \phi  +\nabla^2 \phi &=& - \pa V/\pa \phi = - 
 V'(\phi) - g^2\chi^2 \phi  
\dlabel{unpertphi}\\
\ddot \chi + 3 H\dot\chi +\nabla^2\chi
&=& -\pa V/\pa \chi = -\( -m^2 + g^2\phi^2 \) \chi
\dlabel{unpertchi} , \eea
and the spatially-averaged energy density and pressure are
\bea
\rho(t)&\equiv& \vev{\rho} = \vev{V(\phi,\chi)} + \vev{\dot\phi^2} + 
\vev{\dot\chi^2} \dlabel{energydensity}\\
P(t) &\equiv & \vev{P} = - 
\vev{V(\phi,\chi)}  + \vev{\dot\phi^2} + \vev{\dot\chi^2}  \dlabel{pressure}
. \eea
{}By virtue of the field equations
the energy  continuity equation is satisfied:
\be
\dot\rho(t) = - 3 H(t)  (\rho(t)+ P(t) )
. \dlabel{econt} \ee 

\eqst{unpertphi}{pressure} hold if $\phi$ and $\chi$ are real fields
(with canonical kinetic terms). In realistic models they are at least the 
moduli of complex fields. More generally
they  correspond to directions in 
a field space that provides a representation
 of some non-Abelian symmetry group (the GUT symmetry, for the waterfall field
of GUT inflation).  That introduces some
trivial modifications, namely 
numerical factors in front of \eqs{energydensity}{pressure},
and further factors when it comes to the generation of $\chi$ from 
the vacuum fluctuation.  For the waterfall field it 
also introduces the non-trivial and essential fact that domain walls will
not form when the waterfall spontaneously breaks the symmetry
$\chi\to-\chi$, which would be fatal to the cosmology. Instead there will
be at most cosmic strings, which are harmless if the inflation scale is
not too high. For clarity we pretend that $\phi$ and $\chi$ are real fields.

The  potential \eqreff{fullpot} was proposed 
 in \cite{andreihybrid}, with $V(\phi)=m_\phi^2 \phi^2/2$. 
But if one demands that the curvature perturbation on cosmological scales
is dominated by $\zeta_\phi$,  observation 
requires $V''(\phi)<0$ while those scales 
leave the horizon. Many forms of $V(\phi)$ have been proposed which 
satisfy that requirement \cite{al,book}, and we will 
not assume any particular form.

\section{Generating the tachyonic mass}
\dlabel{stach}

To get firm estimates, we  make the following assumptions;
(i) the  waterfall  starts during slow-roll inflation,\footnote
{The alternative is for slow roll to end
at $\phi>m/g$.
Then $\phi$ will start to oscillate about
its minimum with an amplitude that is (at least) Hubble-damped. In that case
inflation continues until the amplitude of the oscillation falls below
$m/g$.}
(ii) $\phi$ and $\dot\phi$ have negligible variation during the waterfall,
(iii)  the waterfall takes much less than a Hubble time.

Since $H\equiv \dot a/a$  changes little in a Hubble time, we can take it to 
 have a  constant value
denoted simply by $H$. Except insofar as it generates $H$, we also ignore
the variation of $a$ during the waterfall, setting it equal to 1.
We set  $t=0$ at the beginning of the waterfall and use a
dimensionless time $\tau\equiv \mu t$. Then we write 
\be
m^2(\phi(t)) = -\mu^3 t,\qquad \mu^3 \equiv   -2g^2\phi\dot\phi
\simeq -2gm\dot\phi
, \dlabel{ourapprox} \ee
with $\mu$ taken to be constant.  

To estimate $\dot\phi$ we use  \eq{calpzetaphi}, evaluated for the 
wavenumber $k\sub{end}=H$ that corresponds to the horizon at the end
of inflation. 
On  cosmological scales, observation  gives for the total curvature 
perturbation $\calpz\half(k) =5\times 10^{-5}$.
Let us define a number $f$ by
\be  5\times 10^{-5} f \equiv  \calp_{\zeta_\phi}\half(k\sub{end}) 
. \ee
Inflation models are typically constructed so that $\zeta_\phi$
accounts for $\zeta$ on cosmological scales \cite{al,book}. 
They also typically 
make $\calp_{\zeta_\phi}(k)$ almost scale-independent and then
$f=1$, but it's possible to choose the potential so that
$\calp_{\zeta_\phi}(k)$ is strongly increasing \cite{klm}. In the latter case
one can have $\calp_{\zeta_\phi}\half(k\sub{end})\sim 10\mone$ 
corresponding to $f\sim 10^3$. As  mentioned in the Introduction, 
that will  give  significant black hole contribution and
 constrain the parameters of the potential.
Finally, there is the possibility that $\zeta_\phi$ is negligible,
$\zeta$ being generated after inflation by a curvaton-type mechanism
\cite{curvaton}; then  $f\ll 1$.

Using \eqs{calpzetaphi}{ourapprox} we have
\be 
\( \frac \mu m \)^3  \simeq 3\times   10^3 \frac g f \( \frac H m \)^2
\dlabel{ratios} . \ee

Our  assumption  $Ht\sub{vev}\ll 1$ 
  corresponds to
\be
Ht\svev \equiv (H/\mu) \tau\svev \ll 1
.\dlabel{secondcon}\ee
What about the consistency of the assumption that $\mu$ is constant?
{}From \eq{ourapprox}, we see that this is equivalent to  $\dot\phi$ and
$\phi$ both having negligible variation (barring an unlikely cancellation).
Let us assume for the moment that the last term of \eq{unpertphi} remains
negligible throughout the waterfall, so that slow-roll inflation 
(\eq{hofv}) continues to hold. Then $\dot\phi$ has little change
during the waterfall (because it has little change in  Hubble time).
Negligible variation for $\phi$ means  $|\dot\phi| t\svev  \ll m/g$
which is equivalent to 
\be
(\mu/m)^2    \ll \tau\svev\mone
\dlabel{firstcon}. \ee
Now consider the last term of \eq{unpertphi}. 
As we shall see, $\chi^2$ is proportional to a prefactor times
$\exp\(\frac43\tau^{3/2} \)$. If we insert a constant
$\phi$ into the last term of \eq{unpertphi} this  gives  a contribution 
 $\Delta\dot\phi \sim -g^2\phi\chi^2/\mu$. Such a contribution has to be
negligible if $\dot\phi$ is to remain slowly varying, and requiring that
we find
\be
g^2 \ll \lambda (\mu/m)^4 
. \dlabel{secondcon2} \ee

We shall  see that  the condition for  a classical
regime to exist is  $\tau\svev \gg 1$, which means that
\eqs{firstcon}{secondcon} require the hierarchies
\be
H \ll \mu \ll m,\qquad g^2\ll \lambda 
. \ee

\section{Waterfall field}
\dlabel{sgen}

\subsection{Quantum and classical contributions}

Working  with the comoving coordinate $\bfx$ and Fourier components
\be
\chi_\bfk(t) = \int d^3x e^{-i\bfk\cdot\bfx} \chi(\bfx,t)
. \ee
The field equation \eqreff{unpertphi} is 
\be
\ddot \chi_\bfk + 3 H \dot\chi_\bfk  
+ \[ -\mu^3 t + (k/a)^2 \] \chi_\bfk =0
. \dlabel{fullmodefu} \ee
Since we are assuming that the waterfall takes much less than a Hubble time,
we can set $H=0$ in this equation. Going to the dimensionless time
 $\tau\equiv \mu t$ it becomes
\be
\frac{d^2 \chi_\bfk(\tau)}{d\tau^2}
= x(\tau,k) \chi_\bfk(\tau),\qquad x\equiv   \tau-   k^2/\mu^2
. \dlabel{modefu}\ee
The solutions of \eq{modefu} are the Airy functions $\Ai(x)$ and $\Bi(x)$. 

In the  quantum theory, $\chi_\bfk$ becomes an operator $\hat\chi_\bfk$.
Working in the Heisenberg picture we write
\bea
\hat \chi_\bfk(\tau) &=& \chi_k(\tau) \hat a_\bfk + 
\chi_k^*(\tau) \hat a_{-\bfk} \\
\[ \hat a_\bfk, \hat a_\bfp \] &=& (2\pi)^3 \delta^3(\bfk-\bfp)
. \eea  
The mode function $\chi_k$ satisfies \eq{modefu}
and its Wronskian is  normalized to $-1$.  We choose
\be
\chi_k =\sqrt{\pi/2\mu} \[ \Bi(x) + i \Ai(x) \]
. \dlabel{chik1} \ee

To understand this choice  we assume that our 
approximations work back to a time
$\tau\sub{initial}\ll -1$. Then there is for all $k$ a  
regime $x\ll -1$, in which 
\be
\chi_k(\tau) = (2\mu)\mhalf |x|\mquarter e^{-i\pi/4} 
e^{-\frac23 i |x|\threehalf}
. \dlabel{chik2} \ee
In this regime,  $x$ is slowly varying ($|dx/d\tau|\ll |x|$), and
$\chi_\bfk$ describes particles with momentum $\bfk$  and 
mass-squared  $m^2(\phi(t))$. We choose the 
 state vector as the  vacuum, such that $\hat a_\bfk|>=0|>$. A 
significant occupation
number is excluded
since the resulting positive pressure would spoil inflation 
\cite{book}.\footnote
{There is still the  problem with the vacuum  state, that we discuss later.}

Using the mode function \eqreff{chik1} we can work out the two-point correlator of $\hat \chi_\bfk$:
\bea
\vev{\hat \chi_\bfk(\tau)\hat \chi_\bfp(\tau)} &=& (2\pi)^3 \delta^3(\bfk+\bfp) 
\calp_\chi(k,\tau)  \dlabel{vevchik}  \\
\calp_\chi(k,\tau) &\equiv& \frac{k^3}{2\pi^2} |\chi_k(\tau)|^2 
.\eea
The spectrum $\calp_\chi$ is independent of the direction of $\bfk$ because the
vacuum is invariant under rotations.  
This gives  the  expectation value of $\hat \chi^2$
\be
\vev{\hat\chi^2(\tau)} = \int^\infty_0 \frac{dk}k \calp_\chi(k,\tau) \dlabel{vevchix}
, \ee
and the expectation values of $\hat\rho_\chi$ and $\hat P_\chi$,
\bea
\vev{\hat \rho_\chi(t)} &=& \frac1{4\pi^2} \int^\infty_0 dk k^2 \[ -\mu^2\tau |\chi_k|^2
+ |\dot\chi_k|^2 + k^2 |\chi_k|^2 \] \dlabel{unpertrho} \\
\vev {\hat P_\chi(t)}  &=& \frac1{4\pi^2} \int^\infty_0 dk k^2 
\[\mu^2\tau |\chi_k|^2
+ |\dot\chi_k|^2 - \frac13 k^2 |\chi_k|^2 \] \dlabel{unpertp}
. \eea
They  are  independent of $\bfx$ because the vacuum is invariant under translations. 

We are going to see that the final value $\tau\svev$ is much bigger than 1.
For any $\tau\gg 1$, there exists a  regime 
  $x\equiv \tau-(k/\mu)^2 \gg 1$ in which 
\be
\chi_k(\tau) \simeq  (2\mu)\mhalf x\mquarter e^{\frac23 x^{3/2}},\qquad
\dot\chi_k \simeq \mu \sqrt x  \chi_k
,  \dlabel{latemode} \ee
the errors vanishing in the limit $x\to\infty$.
Since the phase of $\chi_k$ is now constant, 
 $\hat \chi_\bfk(\tau)= \chi_k(\tau) ( \hat a_\bfk + \hat a_{-\bfk})$ 
to high accuracy. As a result, $\hat\chi_\bfk$ is  a constant operator times a c-number,
which means
 that $\chi_\bfk$ is a classical quantity
in the WKB sense.
By this, we mean  that a measurement of $\chi_\bfk$ at a given time will give a state
that corresponds to a definite value $\chi_\bfk$ at all future times \cite{book}.

After such a  measurement,  $\chi_\bfk$ is  a classical
field, and  the vev  $\vev{}$  refers 
 to the ensemble of universes corresponding
to different outcomes of the measurement. 
(We have nothing to say about the cosmic
Schro\"dinger's Cat problem that now presents itself.)
The classical perturbation $\chi(\bfx,\tau)$, built from the 
classical modes, is  statistically homogeneous and isotropic. 
Its mean-square is given by \eq{vevchix}, keeping only the classical modes.
Up to cosmic variance, the vev in \eq{vevchik} can be
regarded as an average over a cell $d^3 k$ in our universe, and
the vev in \eq{vevchix} can be regarded as a spatial average in our universe.

\subsection{Dropping the quantum  contribution}

Before continuing, we comment on the procedure of dropping the quantum
regime. 
We saw earlier that $\chi_\bfk$ has a particle interpretation
 in the the regime $x\ll -1$.
Consider   the part of this regime in which the particles
have negligible mass, $|m^2(\phi)|\ll k^2$. 
The  vacuum fluctuation of a massless free scalar field 
gives infinite contributions to the mean-square field, the 
energy density and the pressure. The usual procedure for avoiding the infinity
is to drop the vacuum fluctuation. As we now explain, 
that raises some questions
especially when the quantum regime is accompanied by a classical 
regime as in the present 
case.
Cutting out the contributions above  some momentum  $\Lambda\sub{UV}$, 
\eqs{unpertrho}{unpertp} give in the limit of large $\Lambda\sub{UV}$
the remaining contributions:
\bea
 \chi^2\sub{vfuc} &=&  \Lambda\sub{UV}^2/6\pi^2 \\
 \rho\sub{vfluc} &=& 3 P\sub{vfluc}  = \Lambda\sub{UV}^4/16\pi^2
. \dlabel{rhopvac} \eea
Following \cite{weinberg}, this 
 constant energy density is widely regarded as a contribution to the 
cosmological constant.
That is not a viable interpretation though \cite{dewitt}
because the cosmological constant  is Lorentz-invariant, hence of the form 
$T_\mn^\Lambda=-\eta_\mn \rho_\Lambda$.
This makes $P_\Lambda=-\rho_\Lambda$ in contradiction with \eq{rhopvac}.

The violation of Lorentz invariance by the vacuum fluctuation may 
seem surprising, since the
vacuum is Lorentz invariant. 
The violation comes of course from the  momentum cutoff, 
which
breaks Lorentz invariance.

In this discussion we have been setting $a=1$.
To understand the relation $\rho\sub{vfluc} = 3P\sub{vfluc}$ we 
need to restore $a(t)$, so that 
\eq{rhopvac} becomes \cite{bdbook} 
\be
\rho\sub{vfluc}  = 3 P\sub{vfluc} 
= \frac1{16\pi^2} \( \frac {\Lambda\sub{UV} }{ a(t) } \)^4
. \dlabel{rhopvac2} \ee
With $\Lambda\sub{UV}$ time-independent, 
this satisfies the energy continuity equation \eqreff{econt}.
In fact, each $k$ mode satisfies the energy continuity equation, because
its action is invariant under the  conformal time translation $d\eta=dt/a$.
Imposing instead a time-independent physical cutoff 
$\Lambda\sub{UV}\su{physical}=a(t)\Lambda\sub{UV}(t)$
 makes $\rho\sub{vac}$
{\em and} $P\sub{vac}$ time-independent. The  energy continuity equation
is now violated, because the physical cutoff breaks the invariance of
the action under   the conformal time translation.

Instead of a momentum cutoff one might invoke a regularisation that preserves
the Lorentz invariance, like Pauli-Villars \cite{zeldovich2}
of  dimensional regularization \cite{akhmedov}. The latter gives
$\rho\sub{vfcluc}=P\sub{vfluc}=0$ for the massless case, and for mass $m$
it gives 
\be
\rho\sub{vfluc} = - P\sub{vfluc} = - \frac{m^4}{64\pi^2}
\[ \ln \( \frac {\Lambda^2\sub{UV} }{ m^2 } \) + \frac32 \]
. \ee
It is not clear what that implies for the waterfall field, whose 
mass-squared is negative (tachyonic) during the waterfall.

We are going to keep only the classical regime $x\gg 1$, which means that
we drop not just the massless regime but 
the entire regime $x\lsim 1$.
The procedure of keeping only the classical regime
 is the usual one in cosmology \cite{book}.
It is applied not only  to the waterfall field, but also to 
scalar fields that may be
created after inflation by preheating mechanisms, and to 
light scalar fields during inflation.
In this last case though, alternative
procedures have been advocated \cite{bdbook,tomsbook,parker,maggiore}
 that we should mention here.

The  light fields during inflation (taken to be almost exponential) 
are  essentially defined as those  that have  the canonical kinetic term,  and
are practically free and massless in all modes. (To be more precise, their
potential is practically quadratic, the effective masses-squared of their perturbations
are much less than $H^2$ in magnitude, and their rate of change is 
given by the slow-roll approximation  $\dot\phi =-V'/3H$.)
A canonically-normalized 
inflaton is one example, and there may be others including
a curvaton-type field that generates a scale-independent contribution to
$\zeta$ after inflation. For each mode, there is a quantum regime
$k\gsim aH$, which for sufficiently small $k$ gives way to  
 a classical regime $k\ll aH$. The transition between the two regimes
is roughly 
the epoch of horizon exit $k=aH$. The usual procedure is to keep all of the
classical modes  and to drop all of the rest. 

The alternative procedures  make no 
distinction between quantum and classical modes. They drop part of the
classical contribution to the energy density, while keeping part of the 
quantum contribution.  This invalidates 
 the  standard result
that a light field perturbation is gaussian with spectrum 
(soon after  horizon) $(H/2\pi)^2$. The resulting modification of 
the standard formula $\calp_{\zeta_\phi}(k)
=(H^2/2\pi\dot\phi)^2$ 
has been worked out for one proposal  \cite{tomsbook,parker}.

\section{Classical waterfall field}
\dlabel{sclass}

\subsection{Spatially averaged quantities}

We are dropping modes with  $k>k\sub{max}(\tau)$,
where the cutoff satisfies
\be
x(k\sub{max})\equiv \tau - (k\sub{max}/\mu)^2 \gg 1  
. \dlabel{xmin} \ee
The exponential
amplification of $\chi(\tau)$ in the classical regime will  make the 
results insensitive to the precise choice of the cutoff.

The spectrum of the classical field vanishes for
$k>k\sub{max}$, and for smaller $k$ is given by
\be
\calp_\chi(k,\tau) = \frac{k^3}{4\pi^2\mu} x\mhalf e^{\frac43 x^{3/2} },\qquad
(k<k\sub{max}(\tau))
. \dlabel{classspec} \ee
At fixed $\tau$, the maximum of $\calp_\chi(k)/k$  is at $k\sub{peak}$ given
by
\be (k\sub{peak}/\mu)^2 \simeq \frac12  \tau\mhalf
\ll 1
. \ee
Modes with $k\gsim \mu$ are negligible which means that the quantum
to classical transition is practically complete when $\tau$ first becomes
much bigger than 1, ie.\  as soon as the classical era begins. 

The  mean-square waterfall  field is
\be
\chi^2(\tau) \equiv \vev{\chi^2(\tau)} \simeq \frac1{4\pi^2\mu} 
\int^\mu_0 x\mhalf e^{\frac43 x^{3/2}} k^2 dk
. \dlabel{vevchisq} \ee
Using  $w\equiv (2/3)\sqrt\tau(\tau-x)$ as the integration variable,
with the identity  
\be
\int^\infty_0 dw \sqrt w e^{-w} = \sqrt\pi
, \ee
we find
\be
\chi^2(\tau)  \simeq \frac{ \sqrt{\pi/2} }{16\pi^2} 
 \mu^2 \tau^{-5/4} e^{\frac43 \tau^{3/2} }
, \dlabel{classvev} \ee
the error vanishing in the limit $\tau\to\infty$.

Since the dominant modes of $\chi$ have $k\ll \mu$,
its spatial gradient is negligible (compared with its time
derivative). Ignoring the spatial gradient terms in
in \eqs{unpertrho}{unpertp} we have
\bea
\rho_\chi(\tau) &\simeq & -\frac12\mu^2\tau \chi^2(\tau) 
+\frac12 \dot \chi^2(\tau) \simeq 0 \dlabel{rhoapprox}  \\
P_\chi(\tau ) &\simeq & \frac12\mu^2\tau \chi^2(\tau) 
+\frac12 \dot \chi^2(\tau)  \simeq \mu^2 \tau  \chi^2(\tau) 
\simeq \dot \chi^2(\tau)
 \dlabel{papprox} 
. \eea
When the waterfall ends, 
$P_\chi(\tau) \sim \lambda\mone \mu^2 m^2$. This is
 much less than the  total $P\simeq - V_0 \simeq \lambda\mone m^4$,
as is required for the consistency of our assumptions.

{}From \eq{hoverm} the waterfall ends at
\be
\tau\svev \sim  \( \ln(\chi_0/\mu) \)^{2/3} 
\sim \( \ln(m/\lambda\half\mu) \)^{2/3} \gg 1
. \dlabel{tauvev} \ee

To get an upper bound on $\tau\svev$ we use $3\mpl^2H^2
\simeq V_0 = \chi_0^2m^2/2$ and $H\ll\mu\ll m$, which give
\be
\tau\svev \lsim \(\ln (\mpl/H) \)^{2/3}
. \dlabel{upperontau}\ee
The inflationary Hubble parameter $H$
 has an upper bound coming from the absence of an observed tensor 
perturbation: $H\lsim 10\mfour \mpl$,  which if saturated 
would give $\tau\svev \lsim 6$.
 In most  inflation models $H$ is not many orders of
magnitude smaller (though small enough 
that the tensor perturbation will never be observed).
In principle though, the only bound is 
$\sqrt{\mpl H}> \MeV$ (to allow successfully Big Bang Nucelosynthesis).
  This corresponds to
$H> 10^{-42}\mpl$ or $\ln(\mpl/H) \lsim  90$.
We conclude that  $\tau\sub{vev}$ is 
 probably $\lsim 10$ and certainly less than 60 or so.

\subsection{Pressure perturbation}

Since the spatial gradient of $\chi$ is negligible, \eqs{rhoapprox}{papprox}
hold locally and we have
\be
\delta P_\chi(\bfx,\tau) = \mu^2 \tau \delta[\chi^2(\bfx,t) ]
\dlabel{deltap} . \ee
 The  spectrum of the pressure perturbation is
\be
\calp_{\delta P_\chi} \simeq \mu^4 \tau^2 \calp_{\chi^2}
. \dlabel{pchispec} \ee
Let us estimate  $\calp_{\chi^2}$.

According to our approximations, there is no correlation between
$\chi_\bfk$ except for the two-point correlator \eq{vevchik} and
the disconnected $2^n$ point correlators starting with
\be
\vev{\chi_{\bfk_1}\chi_{\bfk_2}\chi_{\bfk_3}\chi_{\bfk_4} }
=\vev{\chi_{\bfk_1}\chi_{\bfk_2}} \vev{\chi_{\bfk_3}\chi_{\bfk_4}}
+ \mbox{permutations} 
. \dlabel{fourpoint} \ee
This allows us to calculate the  spectrum $\calp_{\delta P_\chi}(k)$, using the 
convolution theorem,
\be
(\chi^2)_\bfk = \frac1{(2\pi)^3} \int d^3k' \chi_{\bfk'} \chi_{\bfk'-\bfk}
.\ee 
One finds \cite{myaxion}
\be
\calp_{\chi^2}(k,\tau) 
=\frac{k^3}{2\pi} \int d^3k' \frac{ \calp_\chi(k',\tau) }{ k'^3 }
 \frac{ \calp_\chi(|\bfk-\bfk'|,\tau) }{ |\bfk -\bfk'|^3 }
. \dlabel{calpchisq}
\ee
We shall use this expression to estimate the contribution to the curvature
perturbation that is generated during the waterfall. For that purpose we
need only $k\ll   H$ while   the dominant modes of $\chi$ have 
$k\gg H$. 
We can therefore set $|\bfk'-\bfk|$ equal to $k'$ on the right hand side
of \eq{calpchisq}, to get
\be
\calp_{\chi^2}(k,\tau) =\frac{k^3}{8\pi^4\mu^2 }
\int^{k\sub{max}(\tau)}_0 dk k^2 x\mone e^{\frac83x^{3/2}}
. \ee
(We are using $k$ also as the integration variable, so that
the definition $x\equiv \tau-(k/\mu)^2$ continues to hold.)
Proceeding as for \eq{vevchisq} we find in the limit $\tau\to\infty$
\be
\calp_{\chi^2}(k,\tau) \simeq \frac{\sqrt \pi}{2^7 \pi^4} k^3\mu
\tau^{-7/4} e^{\frac83\tau^{3/2} }
. \ee

After smoothing on a scale $k$, the
 mean-square  perturbation in $\chi^2$ is of order
$\calp_{\chi^2}(k)$, and the fractional mean-square perturbation is
of order
\be
 \frac { \calp_{\chi^2}(k,\tau) }{ \chi^4(t) } \simeq
4 \( \frac k\mu \)^3 \tau^{3/4}
. \dlabel{fracmsq} \ee
This  will be much less than 1 except perhaps on the very smallest
scales  $k\sim H$. As a result,
 $\delta P_\chi$ can be treated as a first-order order cosmological 
perturbation.

\section{Primordial curvature perturbation $\zeta$}
\dlabel{sprim}

\subsection{Time-dependence of $\zeta$}

To define $\zeta$  one 
smooths the metric and the energy-momentum tensor
on a comoving scale $k\sub{smooth}$, that is outside the horizon, 
yet below the  shortest scale of interest \cite{book}.\footnote
{A function of position is said to be smooth on a given scale $k\sub{smooth}$,
if it has no fourier components with bigger $k$. If a function has such modes, it is smoothed
by  removing them.}
In the gauge with 
the comoving threading and the slicing of uniform
energy density, the spatial metric defines $\zeta$ through the expression
\be
g_{ij}(\bfx,t) \equiv a^2(t)e^{2\zeta(\bfx,t) } \gamma_{ij}(\bfx,t)
. \dlabel{gij} \ee
In this expression, $a$ is the unperturbed scale factor, 
and $\gamma$ has unit determinant. 

The local scale
factor (such that a comoving volume is proportional to $a^3$) is 
$a(\bfx,t)\equiv a(t)\exp[\zeta(\bfx,t)]$. 
 Since the smoothing scale is outside the horizon, no energy can flow on bigger scales
which means that 
the energy continuity equation is satisfied locally:
\be
\dot\rho(t) = - 3\frac{\dot a(\bfx,t) }{a(\bfx,t)} 
\[ \rho(t) + P(t) + \delta P\sub{nad}(\bfx,t) \]
, \ee
where $\dpnad$  is the pressure perturbation on the slicing of uniform density
(non-adiabatic pressure perturbation). We can choose the unperturbed quantities
$P(t)$ and $\rho(t)$ 
so that   the unperturbed energy continuity equation 
\eqreff{econt} is satisfied.
Then  the time-dependence of $\zeta$ is given by
\be
\dot\zeta(\bfx,t) = -\frac{H(t)}{\rho(t) + P(t) + 
\dpnad(\bfx,t)} \dpnad(\bfx,t)
, \ee
where $H\equiv \dot a(t)/a(t)$.
During  any era when $P$ is a unique function of $\rho$, $\dpnad$ vanishes and
$\zeta$ is constant.

To first order in the perturbations, we have
\be
\dot\zeta_\bfk(t) = -\frac{H(t)}{\rho(t) + P(t)} \delta P_\bfk\su{nad}(t)
\dlabel{zetadot} , \ee
and 
\be
\delta P_\bfk\su{nad}(t) = \delta P_\bfk(t) - \frac{\dot P(t)}{\dot \rho(t)} 
\delta\rho_\bfk(t)
= \delta P_\bfk(t) + \frac{\dot P(t)}{3H(\rho(t)+P(t) )} 
\delta \rho_\bfk(t)
, \dlabel{pnad} \ee
where the right hand side can be evaluated in any gauge.

\subsection{The waterfall contribution to $\zeta$}

We define the waterfall contribution  to $\zeta$ as
\be
\zeta_{\chi\bfk} \equiv 
\zeta_\bfk(\tau\svev) - \zeta_\bfk(\tau_*)
= -\frac H\mu \int^{\tau\svev}_{\tau_*} d\tau 
\frac{ \delta P_\bfk\su{nad}(\tau) }{ \rho(t) + P(t)}
, \ee
with $1\ll \tau_*\ll \tau\svev$.
 We are going to see that the integral is dominated
by the upper limit for any choice $\tau_*$ in this range.

The inflaton contribution to $\delta P\sub{nad}$ vanishes 
 (corresponding to $\dot\zeta_\phi=0$) and the waterfall contribution is
just $\delta P_\chi$. Also,
the consistency condition \eqreff{secondcon2} implies $\dot\phi^2
\gg \dot\chi^2$, which means that $\rho + P \simeq \dot\phi^2$.
Using \eq{deltap} we therefore have
\be
\zeta_{\chi\bfk} \simeq -\frac{H\mu}{\dot\phi^2} \int^{\tau\svev}
_{\tau_*} d\tau \tau [\delta(\chi^2(\tau))]_\bfk
,\ee
where $[\delta(\chi^2(\tau))]_\bfk$ is the Fourier transform of
$\delta[\chi^2(t,\bfx)]$.
Since $\delta[\chi^2(\bfx,\tau)]$ increases like
$\exp(\frac43 \tau^{3/2} )$ up to a prefactor,  we have in the limit
$\tau\svev\to\infty$
\be
\zeta_{\chi\bfk} = -\frac12 \frac{H\mu}{\dot\phi^2} \tau\svev^{1/2}
[\delta(\chi^2(\tau))]_\bfk
, \dlabel{zetachiofchisquared} \ee
which gives
\be
 \calp_{\zeta_\chi}(k) \simeq \frac{36}{\sqrt{2\pi}} \tau\svev^{7/4}
\( \frac  H\mu  \)^7 \( \frac kH \)^3 =
\frac{36}{\sqrt{2\pi}} \tau\svev^{-21/4}
\( H t\svev \)^7 \( \frac kH \)^3
. \dlabel{calpzchi} \ee
The black hole bound is
 $\calp_{\zeta_\chi}(k=H) \lsim 10\mtwo$. This will almost certainly be satisfied by \eq{calpzchi},
because it is derived under our assumptions that include
$H t\svev \ll 1$ and $\tau\svev \gg 1$.

On a bigger scale $k= e^{-N}H$, leaving the horizon $N$ Hubble times before the end
of inflation, the $k^3$ dependence makes
$\calp_{\zeta_\chi} \lsim 10^{-(2+ 1.3N)}$ negligible
compared with the total observed value $\calp_\zeta(k)\sim 10^{-9}$
unless $N(k)< 5$. This is means that $\calp_{\zeta_\chi}$ is negligible
on cosmological scales unless $N_0<20$, which is impossible with any reasonable
post-inflationary cosmology even
 with a
 low inflation scale. We conclude that the black hole constraint on the scale
$k=H$ makes $\calp_{\zeta_\chi}$  negligible on cosmological scales,
with our assumptions.

\section{Conclusion}
\dlabel{sconc}

In this paper we have made some observations 
on the usual procedure of dropping the quantum regime. Keeping only the
classical regime, we went on follow the waterfall field and its contribution
to the pressure perturbation. Then we arrived at an expression for the
contribution to the curvature perturbation that is generated during the
waterfall. To do that, we made simplifying assumptions;  that the waterfall
starts during slow-roll 
 inflation,  that it
takes much less than a Hubble time and that $\phi$ and $\dot\phi$
have negligible variation.  On the other hand, we assumed nothing
about the form of the inflationary potential. Our results depend on
five  parameters; the Hubble parameter $H$, 
the tachyonic mass-squared $m^2$ and self-coupling $\lambda$ of the 
waterfall,  its coupling $g^2$ to the inflaton, and fraction $\sqrt f$
 of the observed 
curvature perturbation that is generated by the inflaton perturbation.
We evaluated, for the first time, the 
region of parameter space in which these assumptions hold.

The fundamental feature of our calculation (shared by the discussion
in \cite{llmw} for preheating after $\phi^2$ chaotic inflation) is that 
the {\em linear} non-adiabatic pressure
perturbation is generated by terms that are {\em quadratic} in the 
field perturbation. 
We  find that the contribution to the curvature perturbation,
generated by the waterfall field
during the waterfall, has a spectrum proportional to $k^3$,
making it  negligible on cosmological scales. We concur with the general
view that such a result will apply to any contribution generated
by a field that is massive during inflation, though we have
no universal  proof for the waterfall let alone for the general case.

\section{Acknowledgments}
The author acknowledges support from
 EU grants MRTN-CT-2006-035863 and UNILHC23792, and 
valuable correspondence with 
D.\ Seery, D.\ Mulryne and D.\ Wands. 



\begin{thebibliography}{999}

\bibitem{book}
 D. H. Lyth and A. R. Liddle, {\it The primordial density perturbation},
Cambridge University Press, 2009; 
{\tt http://astronomy.sussex.ac.uk/~andrewl/PDP/errata.pdf};
{\tt http://astronomy.sussex.ac.uk/~andrewl/PDP/extensions.pdf}.

\bibitem{andreihybrid}
 A.~D.~Linde,
  Phys.\ Lett.\ B {\bf 259} (1991) 38.

\bibitem{myfirstaxion}
 D.~H.~Lyth,
  Phys.\ Lett.\  B {\bf 236} (1990) 408.

\bibitem{axionstring}
  D.~H.~Lyth and E.~D.~Stewart,
  Phys.\ Rev.\  D {\bf 46} (1992) 532.

\bibitem{cllsw}
 E.~J.~Copeland, A.~R.~Liddle, D.~H.~Lyth, E.~D.~Stewart and D.~Wands,
                 Phys.\ Rev.\ D {\bf 49} (1994) 6410.

\bibitem{gutinf}
 G.~R.~Dvali, Q.~Shafi and R.~K.~Schaefer,
  Phys.\ Rev.\ Lett.\  {\bf 73}, 1886 (1994)

\bibitem{supernatural}
  L.~Randall, M.~Soljacic and A.~H.~Guth,
  Nucl.\ Phys.\  B {\bf 472} (1996) 377;

\bibitem{mytev}
D.~H.~Lyth,
`Constraints on TeV-scale hybrid inflation and comments on non-hybrid                         
alternatives,''                                                                                
  Phys.\ Lett.\ B {\bf 466} (1999) 85.

\bibitem{runningmass}
 E.~D.~Stewart,
Phys.\ Lett.\ B {\bf 391} (1997) 34;
E.~D.~Stewart,
Phys.\ Rev.\ D {\bf 56} (1997) 2019.

\bibitem{dbi}
 M.~Alishahiha, E.~Silverstein and D.~Tong,
  Phys.\ Rev.\  D {\bf 70} (2004) 123505.

\bibitem{inverted}
  D.~H.~Lyth and E.~D.~Stewart,
  Phys.\ Rev.\ D {\bf 54}, 7186 (1996).

\bibitem{smoothhybrid}
  E.~D.~Stewart,
  Phys.\ Lett.\ B {\bf 345}, 414 (1995);
 G.~Lazarides and C.~Panagiotakopoulos,
  Phys.\ Rev.\ D {\bf 52} (1995) 559.

\bibitem{al}
 L.~Alabidi and D.~H.~Lyth,
  JCAP {\bf 0605} (2006) 016.

\bibitem{klm}
  K.~Kohri, D.~H.~Lyth and A.~Melchiorri,
  JCAP {\bf 0804} (2008) 038.

\bibitem{curvaton}
D. H. Lyth and D. Wands, Phys. Lett. B {\bf 524}, 5 (2002);
T. Moroi and T. Takahashi, Phys. Lett. B {\bf 522}, 215 (2001),
Erratum-ibid B {\bf 539}, 303 (2002).


\bibitem{weinberg}
 S.~Weinberg,
  Rev.\ Mod.\ Phys.\  {\bf 61} (1989) 1.

\bibitem{dewitt}
B.\ S. DeWitt, Phys.\ Rep.\ {\bf 19} (1975) 295.


\bibitem{bdbook}
 N.~D.~Birrell and P.~C.~W.~Davies,
{\it Quantum Fields In Curved Space},
  Cambridge University Press, 1982.


\bibitem{zeldovich2}
  Y.~B.~Zel'dovich,
  Sov.\ Phys.\ Usp.\  {\bf 11} (1968) 381.


\bibitem{akhmedov}
 E.~K.~Akhmedov,
  arXiv:hep-th/0204048.

\bibitem{tomsbook}
L.~Parker and D. Toms, {\it Quantum Field Theory in Curved Spacetime: Quantized
Fields and Gravity}, Cambridge University Press, 2009.

\bibitem{parker}
  I.~Agullo, J.~Navarro-Salas, G.~J.~Olmo and L.~Parker,
  Phys.\ Rev.\ Lett.\  {\bf 103} (2009) 061301;
  I.~Agullo, J.~Navarro-Salas, G.~J.~Olmo and L.~Parker,
  arXiv:1002.3914 [gr-qc].

\bibitem{maggiore}
 M.~Maggiore,
  arXiv:1004.1782 [Unknown].

\bibitem{myaxion}
  D.~H.~Lyth,
  Phys.\ Rev.\  D {\bf 45} (1992) 3394.

\bibitem{llmw}
  A.~R.~Liddle, D.~H.~Lyth, K.~A.~Malik and D.~Wands,
  Phys.\ Rev.\  D {\bf 61} (2000) 103509
  [arXiv:hep-ph/9912473].


\end{thebibliography}
\end{document}